\documentclass[acmlarge]{acmart}
\AtBeginDocument{%
  \providecommand\BibTeX{{%
    \normalfont B\kern-0.5em{\scshape i\kern-0.25em b}\kern-0.8em\TeX}}}

\acmBooktitle{India HCI 2023,
 November 23--25, 2023, Dehradun, India} 
\acmPrice{15.00}
\acmISBN{978-1-4503-XXXX-X/18/06}

\usepackage{url} % For formatting URLs
\usepackage{hyperref} % For creating clickable links
\usepackage{enumitem} % For customizing enumerate labels
\usepackage{booktabs}
\usepackage{graphicx}
\usepackage{lscape}
\usepackage{wrapfig}

\usepackage{graphicx}
\usepackage{caption}
\usepackage{subcaption}
\usepackage{array}
\usepackage{tabularx}
\usepackage{wrapfig}
\usepackage{subcaption}

\usepackage{xcolor, soul}


\begin{document}

% \title{An Interface Designed for Mental Arithmetic Tasks in a Trier Social Stress Test}
\title{Understanding Stress: A Web Interface for Mental Arithmetic Tasks in a Trier Social Stress Test}

\author{Manjeet Yadav}
\email{manjeet19@iiserb.ac.in}
\author{Nilesh Kumar Sahu}
\email{nilesh21@iiserb.ac.in}
\affiliation{%
  \institution{Indian Institute of Science Education and Research, Bhopal}
  \city{Bhopal}
  \state{Madhya Pradesh}
  \country{India}
  \postcode{462066}
}
\begin{abstract}
  Stress is a dynamic process that reflects the responses of the brain. Traditional methods for measuring stress are often time-consuming and susceptible to recall bias. To address this, we aimed to investigate the changes in heart rate (HR) while individuals engaged in the Trier Social Stress Test (TSST). We designed and created an interface for a TSST, incorporating varying levels of complexity in mental arithmetic problems. Our pilot study reveals that participants' HR increased during the Mental Arithmetic Task phase compared to both the baseline and resting stages, indicating that stress is indeed reflected in an individual's HR.
\end{abstract}


\maketitle

\section{Introduction}
Stress is a state of mental tension or worry arising from challenging situations such as work pressure, changes, etc. It is a natural human response that addresses challenges and threats in our lives. A certain stress level is normal, but excessive stress can adversely affect our body, causing headaches, back pain, shoulder and neck discomfort, and digestive issues \cite{mishra2020evaluating}. Additionally, people may experience difficulty focusing, increased irritability, sleep problems, emotions of sadness or guilt, and changes in appetite when under significant stress \cite{MALIK201731}. Everyone experiences stress at some point in time. However, how individuals respond to it greatly influences their overall well-being. 

Stress is typically assessed through self-report surveys or by measuring the body's cortisol level, the stress hormone \cite{mishra2020evaluating}. Cortisol measurement necessitates frequent sampling of blood, urine, or saliva to assess the levels of this stress hormone. According to a study by Cay et al. \cite{cay2018effect}, cortisol levels can increase up to nine times the normal range during stressful periods. Self-report tools like the Perceived Stress Scale (PSS) and the State-Trait Anxiety Inventory (STAI) consist of questions where individuals rate their responses on a Likert scale. However, measuring cortisol levels involves regular visits to a clinic, which can be inconvenient. On the other hand, self-report methods are susceptible to recall bias, where individuals may not accurately remember or report their stress experiences, leading to potential biases in the data.

Research suggests that monitoring changes in Electrocardiogram (ECG) patterns can effectively identify stress levels in individuals during specific situations \cite{giannakakis2019review,mishra2020evaluating,ciabattoni2017real,MALIK201731}. ECG provides the capability to trace variations in heart rate (HR) that occur due to stress, and by analyzing these ECG signals, it becomes possible to determine whether an individual is experiencing stress in a particular situation or while performing a specific task. The advancements in wearable sensor technology and sensing capabilities like wearable Electrocardiogram (ECG) and Photoplethysmography (PPG) bands have empowered researchers to explore the feasibility of continuously detecting and monitoring stress in various settings such as controlled environment scenarios.

In a related vein, we have utilized wearable ECG technology to investigate whether ECG signals exhibit changes during a modified Tier Social Stress Task (TSST) version. The TSSTs deliberately induce stress in individuals within a controlled research setting. The TSST involves various activities, such as solving mathematical problems, giving speeches or presentations, etc. In this study, we developed a web interface consisting of a Mental Arithmetic Task (MAT) featuring increasing difficulty levels to induce stress in individuals. The research question we aimed to address is ``whether HR can effectively reflect an individual's stress levels when they are engaged in solving MAT''. In order to investigate and answer our research question, we conducted a pilot study in which participants completed the designed MAT while wearing ECG electrodes attached to their bodies. A substantial amount of work has already been completed in the above said domain, which typically involves a significant laboratory setup or the performance of various activities in the presence of other participants. In contrast, our approach is capable of inducing stress through a simple web-based interaction accessible on personal computers and mobile phones. Additionally, we can capture and analyze the effects of the induced stress using ECG/PPG sensors. Our results have also demonstrated that the MAT, when performed, induced stress in the participants.

\section{Study Design}

\begin{figure}[!h]
    \centering
    \includegraphics[width=\textwidth, height=0.2\textheight, keepaspectratio]{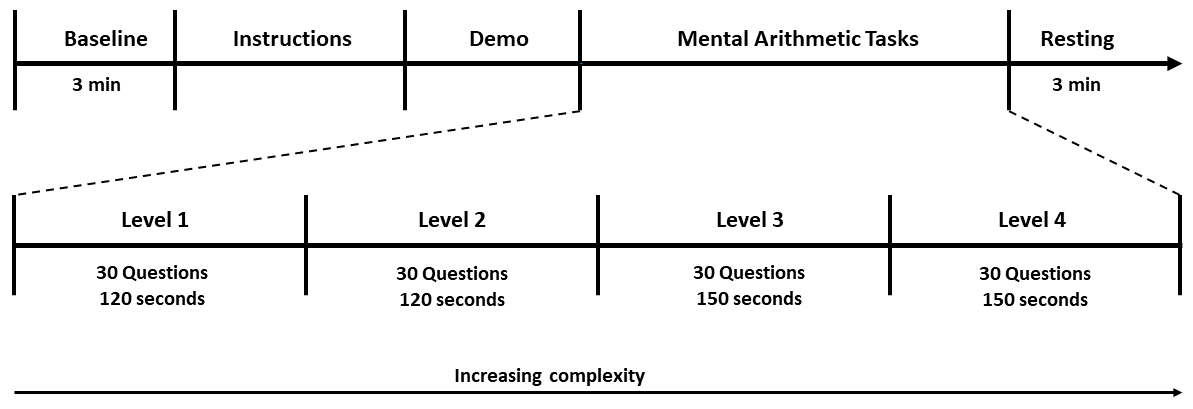}
    \caption{Activity sequence of the designed study}
    \label{fig:study_design}
\end{figure}

\begin{figure}[!ht]
    \centering
    \begin{subfigure}{0.454\textwidth}
        \includegraphics [width=\textwidth]{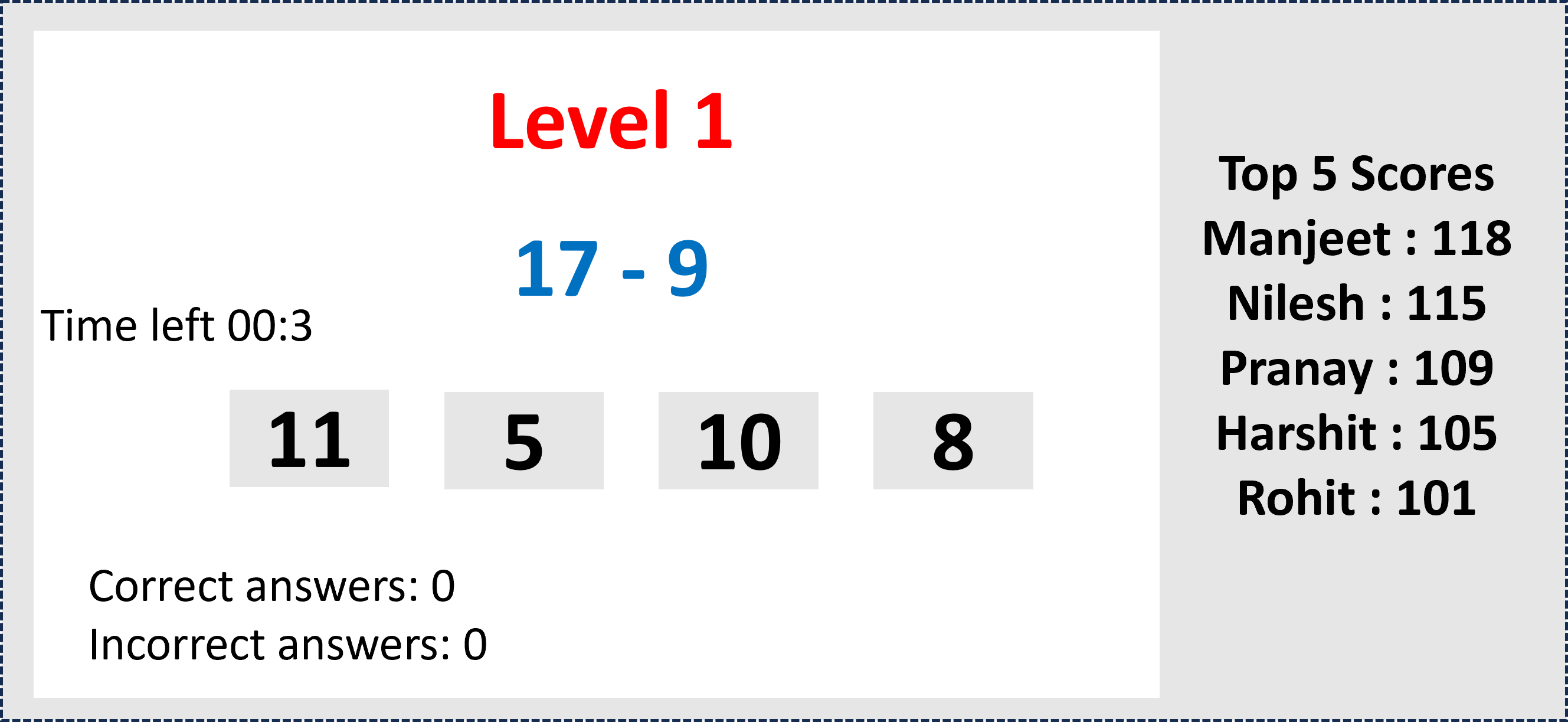}
        \caption{Level 1}
        \label{fig:Level_1}
    \end{subfigure} \hfill
        \begin{subfigure}{0.454\textwidth}
        \includegraphics [width=\textwidth]{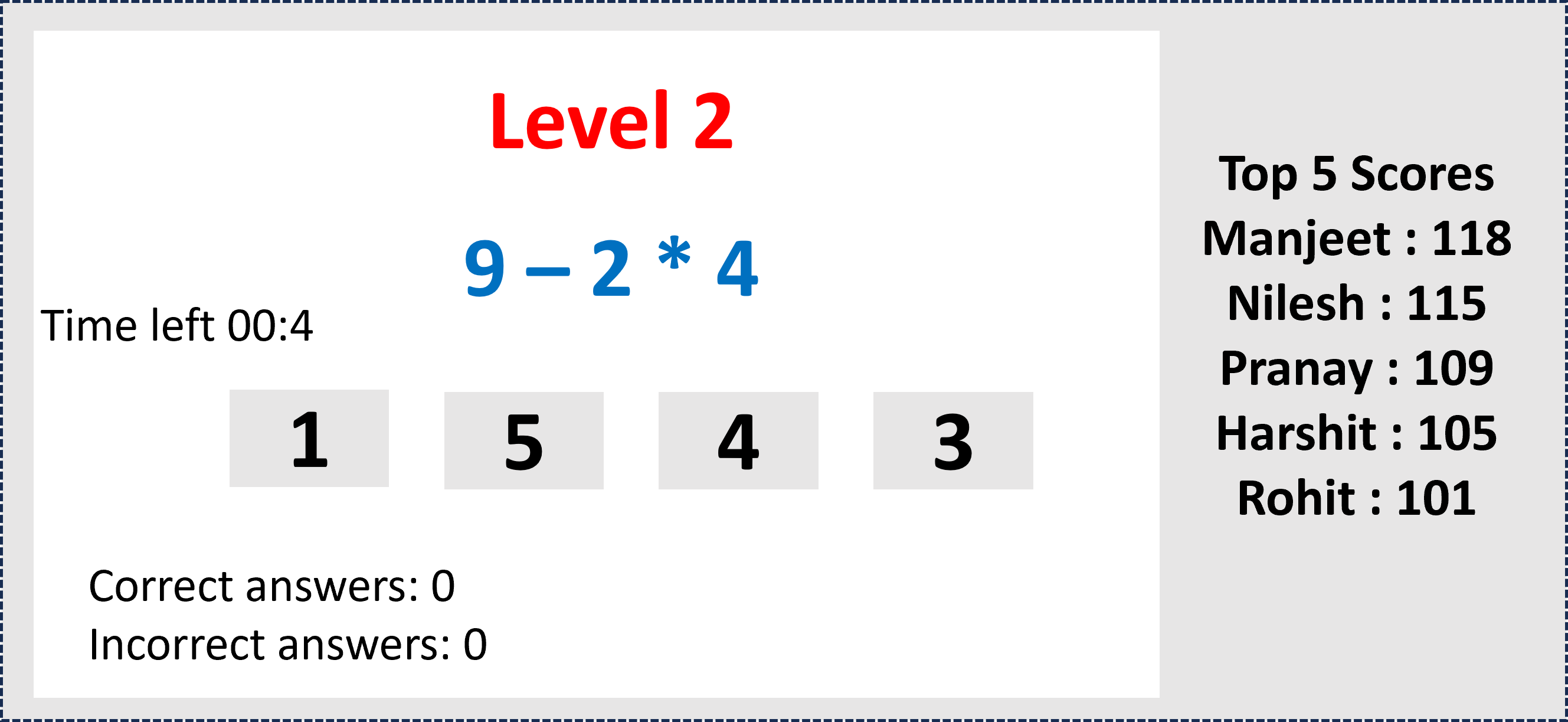}
        \caption{Level 2}
        \label{fig:Level_2}
    \end{subfigure} 
    % \medskip %<- create vertice space
    \begin{subfigure}{0.454\textwidth}

        \includegraphics [width=\textwidth]{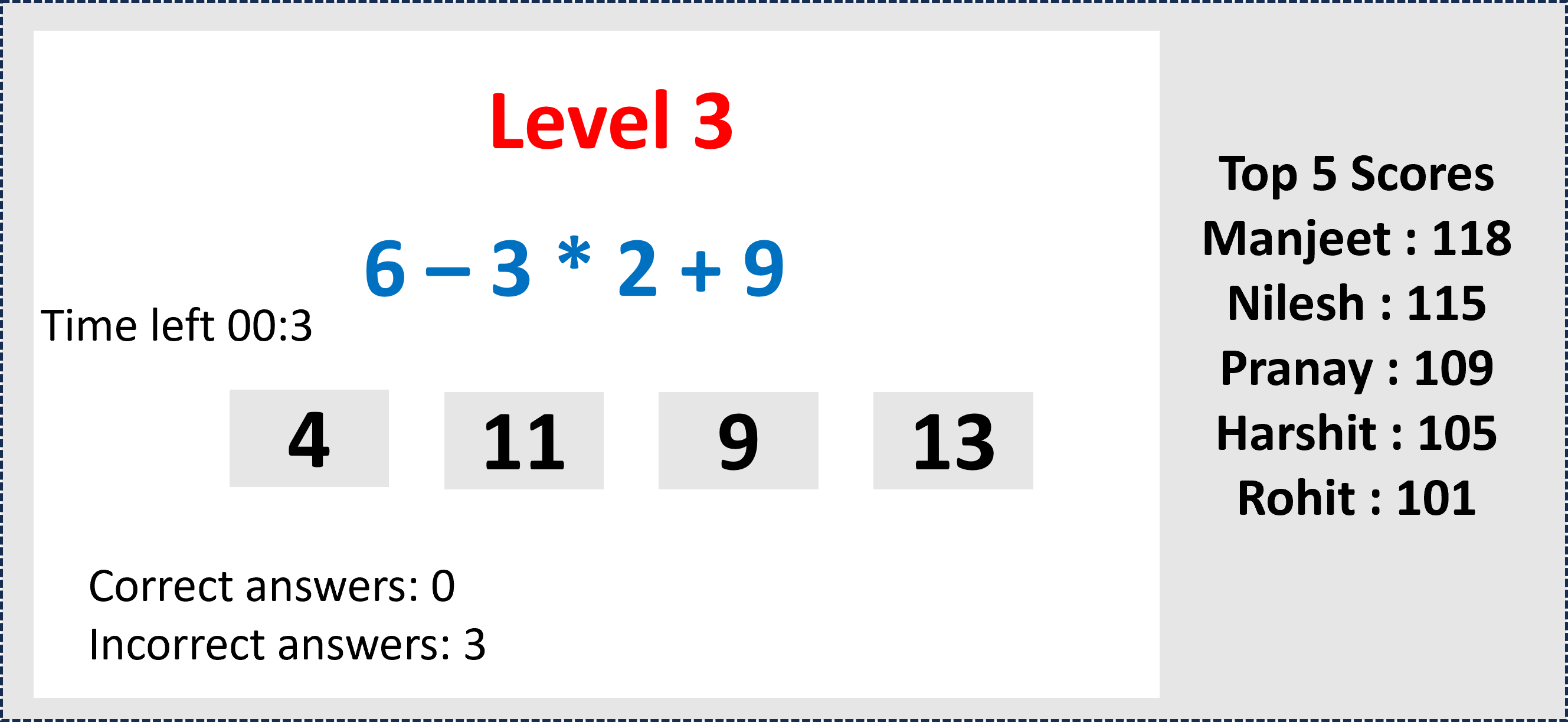}
        \caption{Level 3}
        \label{fig: Level_3}
    \end{subfigure} \hfill
    \begin{subfigure}{0.454\textwidth}
        \includegraphics [width=\textwidth]{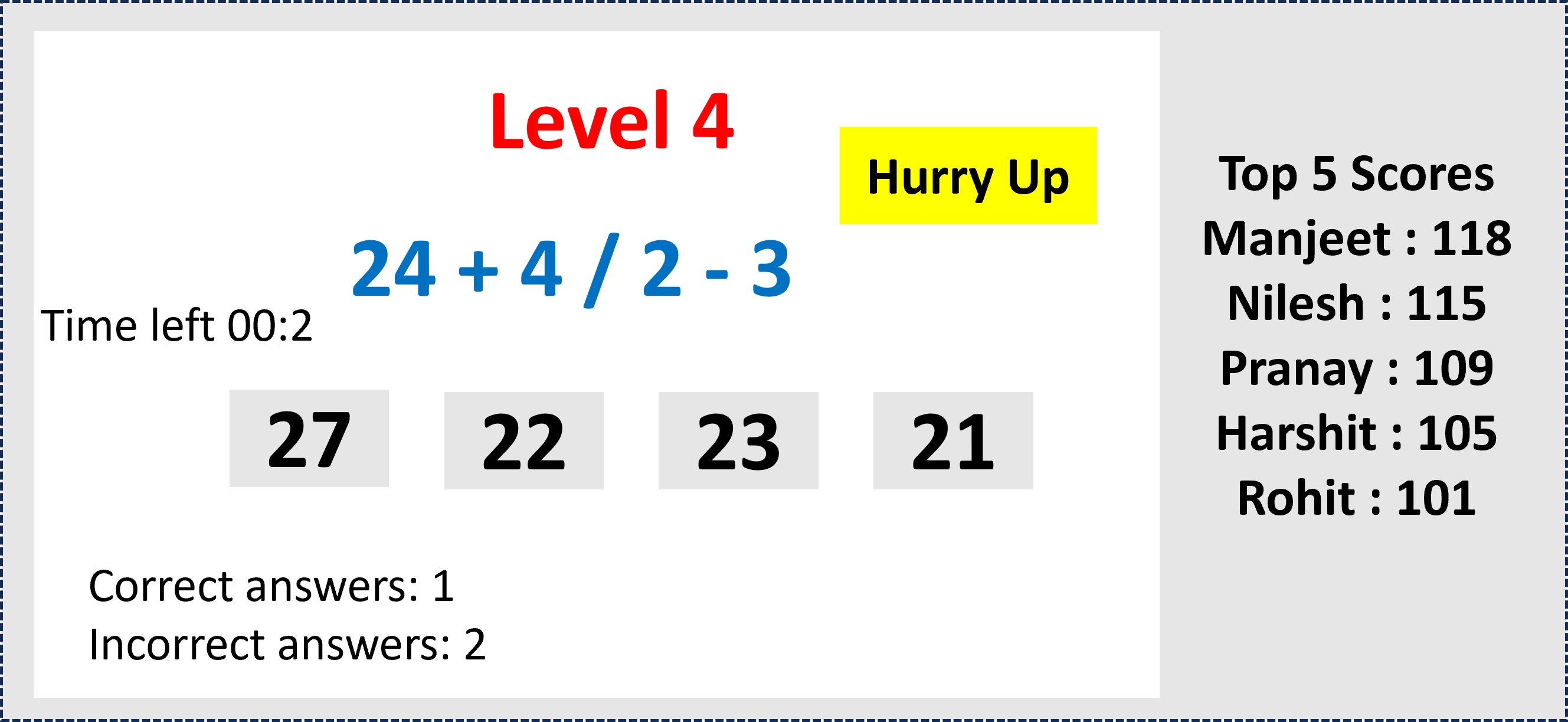}
        \caption{Level 4}
        \label{fig:Level_4}
    \end{subfigure}
    \caption{The web interface of the main working area at different MAT difficulty levels}
    \label{fig:web_interface}
\end{figure}

This section describes the user interface we developed to induce stress, as illustrated in Figure \ref{fig:study_design}. We designed a website \footnote{\url{https://mjyadav7.github.io/MAT_QUIZ.github.io/}} for the MAT using HTML, CSS, and Javascript. 
The website's landing page features a welcoming message inviting users to click an ``agree'' button to provide their consent for participation in the study. Upon agreeing, a 3-minute baseline period begins, accompanied by relaxation music\footnote{\url{https://www.youtube.com/embed/AGkG2BpcFD8?autoplay=1}}. Subsequently, an instruction page appears, offering users an explanation of the MAT's flow. Following the instruction page, a demo trial is presented to give users a brief experience before they embark on the study. The study consists of a quiz with four levels of questions, each level comprising 30 questions. Here are the descriptions of each level, and the web interface is shown in Figure \ref{fig:web_interface}.

\begin{enumerate}
    \item \textbf{\textit{Level 1}} consists of addition and subtraction questions with two numbers such as $9 + 7$. For each question, the user has 4 seconds to answer. So, Level 1 takes 120 seconds ( = 30 questions x 4 seconds/question). 
    %For example, 9+7.

    \item \textbf{\textit{Level 2}} consists of addition, subtraction, and multiplication questions with three numbers such as $8 - 4 * 2$. Again, the user has 4 seconds for each question. So, Level 2 also takes 120 seconds.

    \item \textbf{\textit{Level 3}} is a bit harder. The user has to solve addition, subtraction, and multiplication questions with four numbers such as $7 * 4 + 6 - 21$ and has 5 seconds for each question. So, Level 3 takes 150 seconds.

    \item \textbf{\textit{Level 4}} is the most challenging level. The user solves addition, subtraction, multiplication, and division questions with four numbers such as $72 / 3 + 9 - 5$, and like Level 3, has 5 seconds for each question. So, Level 4 also takes 150 seconds. 

\end{enumerate}

Additionally, the interface includes a scoreboard, displaying dummy names of the top five participants with their scores. Furthermore, to heighten the sense of competitiveness and intensify the induced stress, a ``Hurry Up'' message is displayed on each question when half of the allotted time remains for solving it. For instance, in Level 4, each question has a time limit of 5 seconds, and the ``Hurry Up'' message begins blinking when 2.5 seconds are left to complete the question, as shown in Figure \ref{fig:Level_4}. Between levels of the quiz, participants are allowed to take a short break at their desks but are not permitted to leave the system. After completing the MAT quiz, a resting period begins, during which a relaxation music\footnote{\url{https://www.youtube.com/embed/AGkG2BpcFD8?autoplay=1}} is played.

\section{Pilot Study and Analysis}
We conducted a pilot study involving eight participants from our laboratory. Initially, the participants completed the Perceived Social Stress (PSS) questionnaire to provide a baseline assessment of their general stress levels. Subsequently, a research assistant placed a Shimmer ECG, PPG, and EDA sensors on the participants' bodies. Shimmer ECG, PPG, and EDA are clinical grade sensors and have been used worldwide to gain physiological insights \cite{burns2010shimmer}. Following sensor placements, the participants get themselves engaged in the study.

\clearpage
\noindent{\textbf{Analysis}}

\begin{figure}[!h]
    \centering
    \begin{subfigure}{0.45\textwidth}
        \includegraphics [width=\textwidth,height=0.57\textwidth]{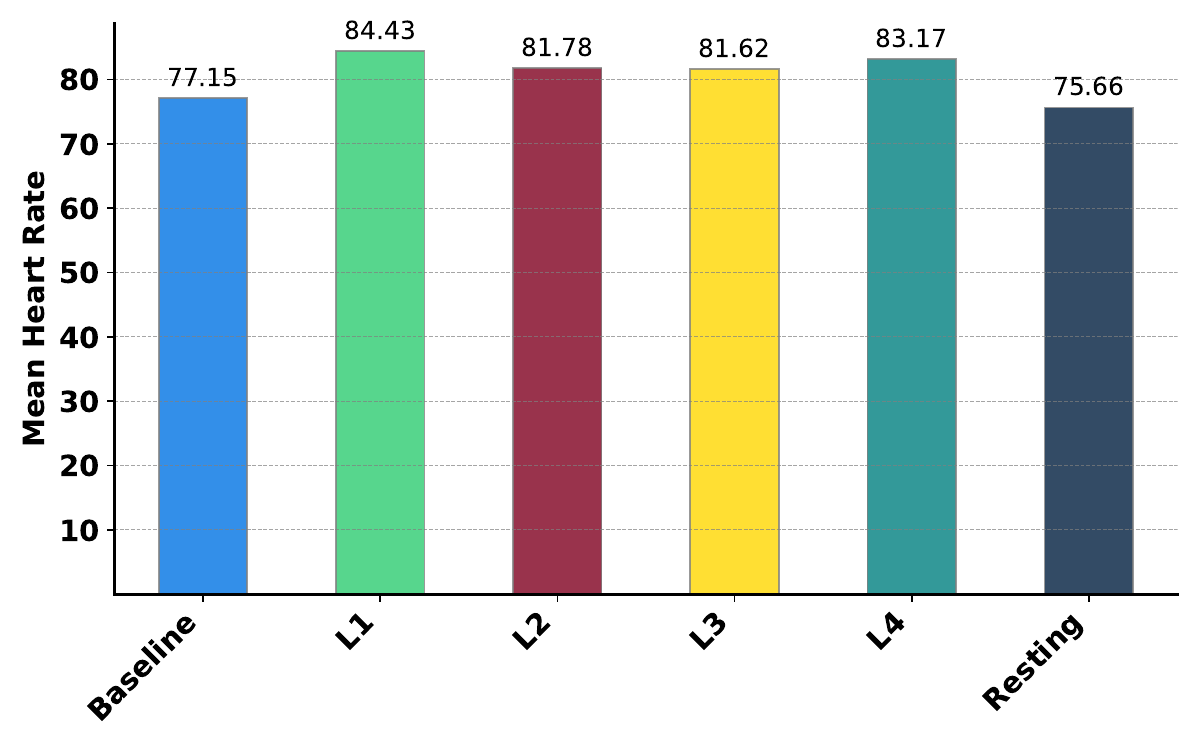}
        \caption{Mean HR}
        \label{fig:mean_hr}
    \end{subfigure} 
    \begin{subfigure}{0.45\textwidth}
        \includegraphics [width=\textwidth,height=0.57\textwidth]{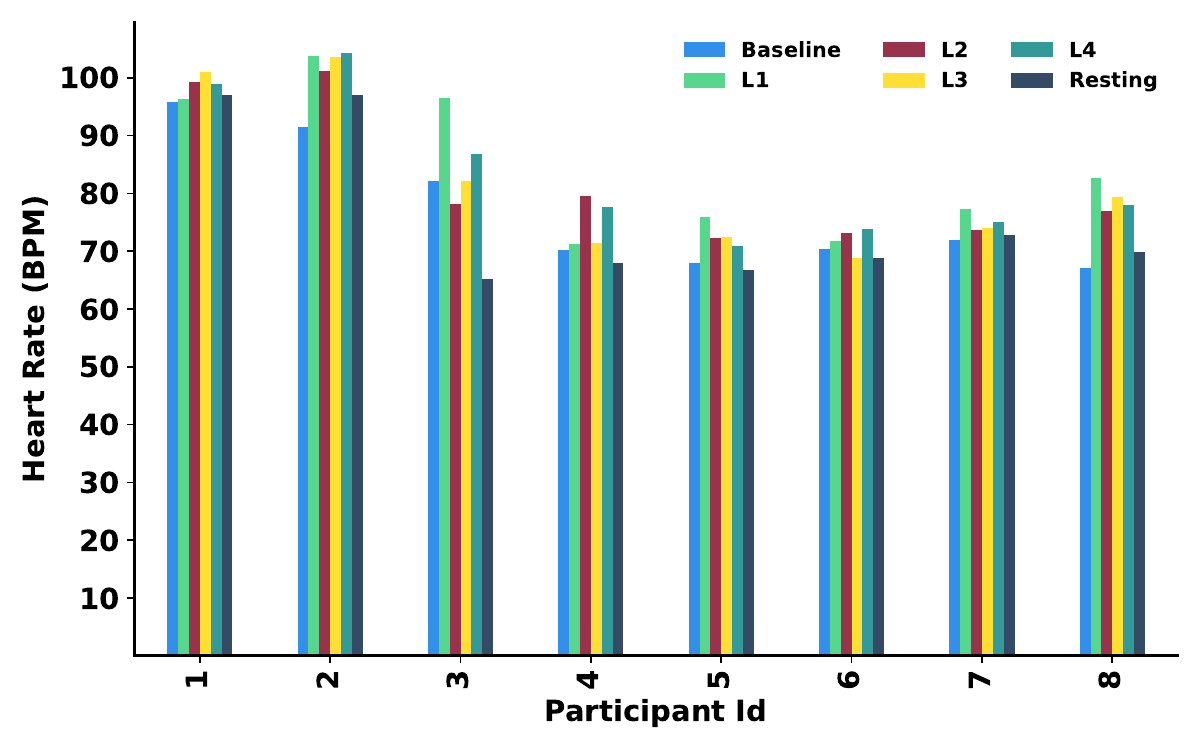}
        \caption{HR of each participants}
        \label{fig:hr_participants}
    \end{subfigure}
    \caption{Heart Rate (HR) in baseline, resting, and each level of quiz.}
\end{figure}

While we collected ECG, PPG, and EDA data, this paper specifically focuses on the analysis of ECG data. Filtering and peak detection algorithms were applied to identify the R peaks in the filtered ECG data. We then utilized Neurokit2 \footnote{\url{https://neuropsychology.github.io/NeuroKit/index.html}}, a Python package, to calculate Heart Rate (HR). Figure \ref{fig:mean_hr} presents the mean HR plot for the baseline, resting, and different quiz levels (i.e., L1, L2, L3, and L4). Additionally, Figure \ref{fig:hr_participants} displays the mean HR data for individual participants during the baseline, resting, and different quiz levels. Notably, we observed the highest mean HR during Level 1 (84.43) followed by Level 2 (83.17), while the lowest mean HR was recorded during the resting phase (75.66), even lower than the baseline (77.15). These findings suggest that participants' HR increases while solving MAT.

\section{Conclusion}
In this research, we investigated the use of heart rate extracted from an ECG signal to understand the changes in stress levels. We developed an interface featuring different levels of Mental Arithmetic Tasks (MAT) to induce stress in participants. Our results reveal that HR was notably higher during the MAT sessions when compared to both the baseline and resting periods.

\begin{acks}
We want to thank our PI Dr. Haroon Rashid Lone and SIRL lab members for their guidance and support. 
\end{acks}

\bibliographystyle{unsrt} 
\bibliography{sample-base}

\end{document}